# Do Large Language Models Have Emotions?

Amit Goldenberg[1,2,3]

James J. Gross[4]

[1] Harvard Business School

[2] Department of Psychology, Harvard University

[3] Harvard University, Digital, Data and Design Institute

[4] Department of Psychology, Stanford University

Corresponding Author: Amit Goldenberg, agoldenberg@hbs.edu

Do LLMs have emotions? A recent paper from Anthropic reports finding internal representations of emotion concepts in Claude Sonnet 4.5, concluding that the LLM has 'functional emotions.' We evaluate this claim against what is known about how emotions actually function in biological systems. We argue that emotions serve two core functions: the context-sensitive interpretation of situations, and the reorganization of processing across multiple systems in response to those interpretations. The Anthropic findings offer partial support for the first function, though the consistent, discrete emotional representations identified in Claude sit uneasily with affective neuroscience findings that human emotion is characterized by variable rather than uniform neural signatures. On the second function, the evidence is mixed: Claude's representations modulate output without producing the dynamic reorganization of attention, decision speed, and motivational state that defines emotion in biological systems. We close by proposing what it would take for an LLM to have emotions.

One of the most intriguing questions today is whether large language models (LLMs) have (or could have) emotions. The question captivates because it cuts to the heart of our deepest fascination with what makes us human. It is a question that deserves a serious answer, and a recent paper from Anthropic represents an intriguing attempt to provide one.

Sofroniew and colleagues investigated the internal representations of Claude Sonnet 4.5, one of Anthropic's flagship LLMs[1]. Their investigation used mechanistic interpretability methods that permit researchers to peer inside the network's activations and identify features that systematically encode specific concepts. Using these methods, Anthropic researchers reported finding 171 internal representations of emotion concepts, abstract features encoding states like curiosity, frustration, and desperation, that track the affective valence of conversational context and causally influence the model's outputs. Crucially, these representations predicted not just what Claude said but also whether it engaged in misaligned behaviors such as sycophancy, blackmail, and reward hacking. By manipulating the representations, researchers altered downstream behavior. For instance, steering the model toward increased activation of the "desperate" vector, while suppressing the "calm" vector, raised the rate of blackmail behavior in a simulated corporate scenario from 22% to 72%, while steering in the opposite direction reduced it to 0%.

Based on these findings, the authors conclude that Claude has "functional emotions," that is, patterns of perception and behavior analogous to those of humans under the influence of an emotion. Their claim is not that Claude feels emotions. The authors sidestep the question of consciousness, asserting instead that Claude has internal states that function like emotions. But what exactly does it mean for an emotion to be functional? The Anthropic paper implicitly treats function as a matter of categorizing situations and shaping outputs, but this captures only a fragment of what emotions actually do. Emotions serve at least two distinct functions: they shape how we read the world, by determining what counts as relevant, threatening, or significant, and they shape how we respond to the world, by reorganizing attention, accelerating decisions, and priming action in ways that go beyond any single cognitive representation. In the sections that follow, we examine each of these functions in turn, asking whether the Anthropic findings provide evidence for either. We then ask what it would take for Claude's functional emotions to actually serve the same functions as human emotions.

## Emotions Shape How We Interpret the World

Emotions help make sense of the world. They do this via appraisal, a process by which the individual evaluates a situation in ways that determine which emotion, if any, is generated in response. Different appraisal profiles, based on whether a situation is seen as novel, relevant to one's concerns, controllable, or caused by oneself or another, give rise to different emotions even in response to identical external stimuli [2,3]. In this respect, the Anthropic findings reveal a point of convergence between LLM processing and human emotion. The model appears to evaluate the contextual features of a conversation and generate internal representations that track emotional significance, somewhat similar to how appraisal operates in humans.

A closer look at the Anthropic findings, however, reveals a key point of difference with the human case. A central insight of appraisal theory is that emotions are flexible and context-dependent. The same event can generate fear in one person, excitement in another, and indifference in a third, depending on how it is appraised. Yet the Anthropic paper treats the discovery of distinct, consistent internal representations for different emotion concepts as evidence of distinct emotional states, implying that each emotion has a reliable neural signature. This misses the flexibility that is central to emotion.

The neuroscience literature reinforces this point. The assumption that specific emotions reliably correspond to distinct, replicable patterns of neural activation has been seriously challenged by a growing body of evidence[4]. A meta-analysis by Lindquist and colleagues[5], which synthesized neuroimaging data from hundreds of studies, found that discrete emotion categories were not associated with consistent or specific activation of distinct brain regions. More recent work by Satpute and colleagues has further demonstrated that emotion representations in the brain are highly distributed, context-dependent, and variable across individuals and instances of the same emotion category [6,7]. Variability, rather than consistency, appears to be the norm.

This has direct implications for the Anthropic findings. If distinct, identifiable internal representations are taken as evidence of distinct emotions, the logic depends on precisely the view of emotion that contemporary affective neuroscience has most forcefully challenged. What

Sofroniew et al. may have identified is better described as abstract semantic representations of emotional concepts rather than the functional states those concepts refer to.

**Emotions Shape How We Respond to the World**

Consider what happens when a person becomes afraid. Attention narrows, decisions accelerate, the body prepares for action. These are not consequences of the emotion. These shifts are part of the emotion. Emotions constitute a package of changes that reorganizes how cognitive resources are allocated across multiple systems simultaneously. Fear narrows attention toward threat-relevant stimuli, speeds threat-related decisions, and biases memory encoding toward emotionally salient material [8]. Anger increases approach motivation and amplifies sensitivity to norm violations. Positive emotions broaden the repertoire of attention and thought, expanding the range of stimuli and associations that are actively processed [9]. These changes unfold with a characteristic temporal signature: more sustained than a reflex, more bounded than a mood, persisting long enough to shape an episode of behavior before dissipating as the eliciting situation resolves [10]. Emotions also orient the organism toward action. Anger motivates approach and confrontation, fear motivates avoidance or flight, and guilt motivates reparative behavior. These action tendencies represent a genuine shift in the motivational state of the organism that changes the probability, speed, and character of its actions in ways that go beyond what any single cognitive representation would predict.

The Anthropic paper focuses on what Claude produces in response to emotionally relevant situations: what it says, what it prefers, whether it behaves sycophantically. But measuring the content of outputs captures only the most visible aspect of an emotional response. What it misses is what emotions do to the processing that generates those outputs. The model does not have a processing architecture that is dynamically reconfigured in response to emotional representations across time and systems. Each forward pass through the network is computationally fixed in its structure. There is no narrowing of attention, no acceleration of certain processing pathways, no sustained shift in how incoming information is weighted. The model's emotional representations modulate what is generated in the same way that any other contextual feature influences generation. This is importantly different from the way emotion reorganizes processing in biological systems, and to describe it as functional emotion conflates the outputs of emotional processing with the process itself.

**So, Does Claude Have Functional Emotions?**

Anthropic uses the term "functional emotions" deliberately, to avoid claiming that Claude actually has emotions while still asserting that something emotion-like is happening. But this framing carries an implicit assumption about what the functions of emotions are, namely detecting emotionally relevant situations and shaping outputs accordingly. This captures something real, but it misses what is arguably the more important functional signature of emotion: the reorganization of processing that emotions produce in biological systems.

We are not troubled by the absence of many characteristics typically associated with human emotion. The lack of facial expressions, physiological arousal, or even conscious feeling is not, in our view, disqualifying. What matters is whether the system shows evidence of two core functions: the context-sensitive interpretation of situations, and the reorganization of processing in response to those interpretations.

On the first, the evidence is mixed. Claude detects emotional significance in flexible ways, but the Anthropic paper's interpretation of this as distinct emotional states sits uneasily with what we know about emotional representation in the brain. In humans, variability rather than consistency is the norm, as the same emotion elicits different patterns of activation across individuals and contexts, because emotions are assembled anew each time through appraisal. A consistent fear vector looks more like a learned semantic representation than a flexible emotional process. On the second function, the evidence is also mixed. Claude's emotional representations do not reorganize processing, narrow attention, accelerate decisions, or shift motivational state. What the Anthropic paper documents is a system that has learned to represent emotional concepts. That is not nothing, but it falls short of functional emotion in any scientifically meaningful sense.

**What Would It Take for an LLM to Have Emotions?**

What evidence would make a more compelling case? We suggest that what is needed is evidence that LLM emotions involve functionally interpretable changes in processing: in attention, processing speed, and response type and likelihood.

Importantly, this would not require emotions to take any specific physical form. The substrate need not be biological for the functional architecture to qualify. As an example, consider the emotions of a collective, also called collective emotions, which are emotions evaluated at the level of multiple individuals rather than a single individual. When a group faces a collective threat and experiences something like collective fear, this state is not instantiated in a single body but is rather distributed across a network of individuals. Yet it has genuine emotional properties: collective attention narrows toward the threat, decision-making accelerates, the group's processing of the world is reorganized around the emotionally relevant stimulus, and these changes persist across the episode[11]. The substrate is social and distributed rather than individual and biological, but the functional signature is present. A similar logic could, in principle, apply to an LLM. One would need to show that the system's emotion-like representations produce sustained, multi-system reorganization of processing, not merely a shift in the probability of the next token, Perhaps one day LLMs will have that. When they do, we’re ready to say that they have emotions.